\documentclass[prd,twocolumn,eqsecnum,nofootinbib]{revtex4}
\usepackage{bm}
\usepackage{amssymb}
\usepackage{graphicx}
\usepackage{epstopdf}
\usepackage{mathrsfs}
\usepackage{amsmath}
\usepackage{color}
\numberwithin{equation}{part}

\begin{document}
\title{
Gravitational waves from a plunge into a nearly extremal Kerr black hole}
\author{Lior M.~Burko$^{1}$ and Gaurav Khanna$^2$}
\affiliation{$^1$ School of Science and Technology, Georgia Gwinnett College, Lawrenceville, Georgia 30043 \\
$^2$ Department of Physics, University of Massachusetts, Dartmouth, Massachusetts  02747}
\date{Draft of \today}
\begin{abstract} 

We study numerically in the time domain the linearized gravitational waves emitted from a plunge into a nearly extremal Kerr black hole by solving the inhomogeneous Teukolsky equation. We consider spinning black holes for which the specific spin angular momentum $a/M=1-\epsilon$, and we consider values of $\epsilon\geq 10^{-6}$. We find an effective transient behavior for the quasi-normal ringdown: the early phase of the quasi-normal ringdown is governed by a decay according to inverse time, with frequency equaling twice the black hole's horizon frequency. The smaller $\epsilon$ the later the transition from this transient inverse time decay to exponential decay. Such sources, if exist, may be interesting potential sources for terrestrial or space borne gravitational wave observatories. 

\end{abstract}

\maketitle

\section{Introduction}

Perturbed black holes settle down to a quiescent state by radiating first like a damped oscillator, with a set of complex-valued frequencies known as the black hole's quasi-normal modes. This exponential quasi-normal ringdown is followed by a power-law decay, the late-time tail part of the radiation. 

In the toy model of a source-free scalar field it has been shown that for a nearly extreme Kerr (NEK) black hole, many weakly damped overtones share the same oscillation (real) frequencies, and the imaginary part of the complex frequencies are nearly equi-spaced in the early part of the ringdown \cite{Yang-2013} (see also \cite{Glampedakis}).  Yang {\it et al} \cite{Yang-2013} showed that this  behavior of the quasi-normal modes leads to an effective decay rate of the superimposed overtones which decays like a power law, specifically inversely in time. It was suggested in \cite{Yang-2013} that the slower decay rate of the superimposed overtones of a NEK black hole in the early part of the ringdown phase could provide a promising strong source for gravitational-wave detectors. Similar behavior of the complex frequencies of the quasi-normal overtones is found also for the gravitational case \cite{dias-2009}, suggesting that a corresponding inverse-time decay rate may be found there too. 

Gralla {\it et al} \cite{Gralla_2016} considered the inspiral of a particle into a more massive NEK black hole in the frequency domain, and found a ``smoking gun'' feature of this scenario, specifically that the frequency of the gravitational radiation coming form the near-horizon region has twice the horizon frequency and the profile is damped exponentially. This property would be evident in the gravitational waves radiating from such a system, and the hallmark of astrophysical NEK black holes. For a NEK black hole the Innermost Stable Circular Orbit (ISCO) is very close to the event horizon, and therefore before the onset of the plunge phase of the motion the outgoing radiation is suppressed because the  particle effectively corotates with the black hole, giving rise to the exponentially damped profile. 
The analysis in  \cite{Gralla_2016} does not extend beyond the ISCO, and therefore does not include the imprint of the eventual plunge on the gravitational waveform. The analysis in  \cite{Gralla_2016}  also did not study in detail the behavior of the quasi-normal modes of the ringdown phase. 

In this paper we combine the two questions, and consider the gravitational radiation emitted from a small particle plunging into a NEK black hole. We find both the ``smoking gun'' frequency first found in  \cite{Gralla_2016}  and the slow decay rate that was first suggested in \cite{Yang-2013} based on the source-free scalar field toy model. Because we are primarily interested in the plunge phase, we do not study the exponential damping that was found in \cite{Gralla_2016}.

The near-horizon geometries of extreme and NEK black holes are diffeomorphic, and display conformal symmetries that allow the mapping of near-horizon circular  trajectories for extreme black holes to plunge trajectories in the near-horizon geometry of NEK black holes \cite{bredberg-2010}.  These conformal symmetries were used in  \cite{Hadar-2014} to find the gravitational waves emitted from the plunge into a NEK black hole, and were given in terms of an infinite sum over the $\ell,m$ modes. 
Our results cannot be directly compared with the results of \cite{Hadar-2014} because we use different plunge trajectories than those used in \cite{Hadar-2014}. It remains an interesting question to directly compare these results. 

Typically, the (linearized) gravitational radiation includes two distinct contributions of differing frequencies: first, the direct radiation of the inspiraling particle, and second, the spacetime response to the particle's lineared perturbations, specifically the quasi-normal modes. In the case of inspiral into a NEK black hole, however, when the particle gets close to the light ring the frequencies of the two sources become comparable, which makes their separation difficult. To predict the gravitational waveforms from NEK black holes that observatories such as LIGO or eLISA may detect one needs therefore to consider a full picture, in which both the particle's direct radiation and the NEK black hole's quasi-normal ringdown radiation are considered simultaneously. 

Black holes are limited by the maximal spin angular momentum they can have. According to the Cosmic Censorship Hypothesis, a black hole of mass $M$ and spin angular momentum $J$ has specific spin angular momentum $a:=cJ/(GM)$ which satisfies $0\le a/M\le 1$. We use below geometrized units in which $G=1=c$. While the Cosmic Censorship Hypothesis does not forbid black holes from being as close to this limit as one wishes (or even at the limit), we know of no dynamical processes that can evolve a gravitating system arbitrarily close to this limit. Realistic processes limit the dynamical formation of fast spinning black holes to $a/M\lesssim 0.998$ \cite{Thorne-74}. (See also \cite{Kesden-2010}.) We are taking here an agnostic viewpoint on the formation of NEK black holes. Specifically, we ask what the gravitational waveforms from a NEK black hole would be, if nature provided a mechanism for  their creation. 

We consider a compact object of mass $\mu$, and a central NEK black hole of mass $M$ and specific spin angular momentum $a$, such that $a/M=1-\epsilon$, where $\epsilon\ll 1$.  Below, we present results scaled for $\mu/M=1$, and for actual low mass ratio values the results can be rescaled with the desired value of $\mu/M$. The orbital evolution is driven by radiation reaction effects during the inspiral phase of the orbital evolution. When the orbital evolution changes to plunge, the radiation reaction effects become less important, because the time scale for radiation reaction is much longer than the dynamical time scale characterizing the plunge phase. We therefore approximate the motion of the particle by geodesic motion for the plunge phase. Comparing our results for the plunge with those obtained from an Effective One Body approach to radiation reaction for large $\epsilon$ ($\epsilon\ge 10^{-4}$) shows consistency. In practice, we present here results for the angular momentum per unit mass of the plunging particle, $L/\mu = 2\, M$ and energy per unit mass, $E/\mu=1$ for motion on the equatorial plane and waves extracted at infinity on the equatorial plane for the azimuthal modes, i.e., modes $m$ such that the multipoles $\ell$ are summed over. We emphasize that the $m=2$ mode does not dominate the radiation, and other azimuthal modes also contribute significantly in the NEK radiation \cite{Gralla-2015}. We have tested other $m$ modes, specifically $m=3$ and $m=4$, with similar results for our main conclusion, specifically the transient inverse-time decay rate of the quasi-normal ringdown. 
For each $m$ mode the greatest contribution to the radiation comes from the $\ell=m$ mode, consistently with the results of \cite{Gralla-2015}.


This paper is organized as follows. In Section \ref{sec2} we describe in detail the numerical technology used including new elements introduced to the inhomogeneous Teukolsky solver that allow us to efficiently answer the interesting questions raised here. In Section \ref{sec3} we present our numerical results for the frequency, amplitude, and energy flux at infinity. We discuss in Section \ref{sec4} the possibility of detection of a gravitational wave signal from a NEK black hole source.

\section{Numerical technology}\label{sec2}

In order to compute the gravitational-wave signal from a small object 
plunging into a NEK black hole, we use point-particle perturbation theory  
for Kerr black hole spacetime. The mathematical formulation in this context is 
the Teukolsky master equation with a particle source-term, which describes 
scalar, vector and tensor field perturbations in the space-time of a 
rotating black hole~\cite{Teuk:1972}
\begin{eqnarray}
\label{teuk0}
&&
-\left[\frac{(r^2 + a^2)^2 }{\Delta}-a^2\sin^2\theta\right]
        \partial_{tt}\Psi
-\frac{4 M a r}{\Delta}
        \partial_{t\phi}\Psi \nonumber \\
&&- 2s\left[r-\frac{M(r^2-a^2)}{\Delta}+ia\cos\theta\right]
        \partial_t\Psi\nonumber\\  
&&
+\,\Delta^{-s}\partial_r\left(\Delta^{s+1}\partial_r\Psi\right)
+\frac{1}{\sin\theta}\partial_\theta
\left(\sin\theta\partial_\theta\Psi\right)+\nonumber\\
&& \left[\frac{1}{\sin^2\theta}-\frac{a^2}{\Delta}\right] 
\partial_{\phi\phi}\Psi +\, 2s \left[\frac{a (r-M)}{\Delta} 
+ \frac{i \cos\theta}{\sin^2\theta}\right] \partial_\phi\Psi  \nonumber\\
&&- \left(s^2 \cot^2\theta - s \right) \Psi = -4\pi\left(r^2+a^2\cos^2\theta\right)T   ,
\end{eqnarray}
where 
$\Delta = r^2 - 2 M r + a^2$ and $s$ is the ``spin weight'' 
of the field. The $s = -2$ version of these equation describes the 
radiative degrees of freedom of the gravitational field in the radiation zone, and can be 
simply related to the Weyl curvature scalar as
$\Psi = (r - ia\cos\theta)^4\psi_4$ which in turn can be directly connected 
with the two polarizations of the emitted gravitational wave at null 
infinity via the relation 
\begin{equation}
\psi_4 \to \frac{1}{2}\left(\frac{\partial^2 h_+}{\partial t^2} - i
\frac{\partial^2 h_\times}{\partial t^2}\right)
\label{eq:psi4hphm}
\end{equation}

The point-particle source-term $T$ on the right hand side is related to the 
particle's energy-momentum tensor 
\begin{equation}
T_{\alpha\beta} 
= \mu\, \frac{u_\alpha u_\beta}{\Sigma\, \dot t\,\sin\theta}\,
\delta\left[r - r(t)\right]\,
\delta\left[\theta - \theta(t)\right]\,
\delta\left[\phi - \phi(t)\right]
\label{eq:source_stress}
\end{equation}
expressed in Boyer-Lindquist coordinates, with $\Sigma = r^2 + a^2
\cos^2\theta$, proper-time $\tau$ and $u_\alpha$ being the particle's 4-velocity. 
The source-term $T$ is constructed by projecting the energy-momentum tensor 
above onto the Kinnersley tetrad and then operating upon that with a 
complicated second-order differential operator~\cite{Teuk:1972}. Additional 
details are available in this Ref.~\cite{Sundar:2007}. 

One remark worth making is on the behavior of the source-term $T$ as the 
particle approaches the horizon. Note the presence of $\dot t \equiv dt/d\tau$ 
in the denominator of  Eq.\ (\ref{eq:source_stress}). As the particle 
approaches the horizon, $\dot t \to \infty$ and that results in the source-term 
$T$ diminishing in magnitude rapidly. As a result of this, the Teukolsky 
equation (\ref{teuk0}) smoothly transitions into its homogeneous form, 
connecting the gravitational radiation from the plunge phase to the Kerr 
hole's quasi-normal modes in a very natural way.

Computing the gravitational waveforms from the capture of a small object by a
Kerr black hole, involves a two-step process. First, we generate a full 
trajectory that would be taken by the particle as it spirals into the black 
hole. As pointed out before, for the purposes of this current work, we simply 
use a plunging geodesic with orbital parameters $E/\mu = 1$ and $L/\mu = 2$. Since 
our main interest here is the very late stage of the binary system's 
evolution, i.e., the plunge and ringdown phases, this is a reasonable simplification.
Once the complete trajectory of the smaller object is available the second 
step in the process is initiated. The Teukolsky equation is solved in the 
time-domain~\cite{Sundar:2010,Zenging:2011,McKennon:2012} by feeding the 
trajectory information into the particle source-term on the equation's 
right-hand-side dynamically. This computation directly generates a high-accuracy 
time-domain waveform for any further analysis or study. The time-domain 
approach is used for the waveform computation (as opposed to the more common 
and simpler frequency-domain approach), because it is more appropriate  
during the late-phase of the system's evolution, since it no longer has any 
periodic motion. 

The numerical methodology we use to solve the Teukolsky equation (\ref{teuk0}) 
is the same as the one presented in our earlier work (see Ref.~\cite{Sundar:2007} 
and references therein).  The main steps of the method are as
follows: (i) we first rewrite the Teukolsky equation using suitable 
coordinates (explained in the following paragraph);  (ii) taking
advantage of axisymmetry, we separate the dependence on azimuthal
coordinate $\phi$, thus obtaining a set of (2+1) dimensional
equations;  (iii) we then recast these equations into a first-order,
hyperbolic partial-differential-equation form; and finally (iv) we implement 
a two-step, second-order Lax-Wendroff, time-explicit, finite-difference 
numerical evolution scheme.  The particle-source term on the right-hand-side 
of the Teukolsky equation requires some care for such a finite-difference 
numerical implementation.  Additional details can be found in our earlier
work~\cite{Sundar:2007,Sundar:2010} and the associated references.

Furthermore, two recent advances made to the time-domain portion of the 
computation have made the process of computation of the time-domain waveform 
highly accurate and efficient. First, a compactified hyperboloidal layer was 
added to the outer portion of the computational domain that allows for the 
extraction of the waveform data directly at null infinity~\cite{Zenging:2011} 
and completely eliminates the ``outer boundary problem'' which is usually a 
challenge for all such computations. Secondly, advances made via parallel 
computing, in particular, OpenCL/CUDA-based GPGPU-computing has allowed for 
the possibility of performing very long duration and high-accuracy computations 
within a reasonable time-frame. With these enhancements, numerical errors in 
these computations are typically on the scale of a small fraction of a 
percent~\cite{McKennon:2012}.

One new enhancement that was deemed necessary for this current work is an 
increase in the numerical {\em precision} associated to the floating-point 
computation, beyond the common double-precision. This is because our main 
focus in this work is on quasi-normal modes that tend to decay rapidly 
(exponentially) and errors arising from finite numerical precision can often 
plague those computations, especially in the context of long evolutions. However,
high-precision floating-point arithmetic is exceedingly expensive computationally, 
even on high-end many-core parallel processors like GPUs. Therefore, we made 
use of a  {\em mixed}-precision approach, wherein the computationally 
expensive source-term $T$ computation was performed with the usual double-precision 
accuracy, but the Lax-Wendroff stepping was done in full quadruple-precision. 
This hybrid approach produced reasonably good quality results within a suitable 
time-frame.  

As pointed out above, an important part of our computational model is the particle approaching the horizon, and the associated natural decay of the source term for the Teukolsky equation. Our computational grid is uniform in the $r^*$ radial coordinate which makes it very well suited to resolve the complex physics that occurs near the hole. This is particularly significant in the context of the NEK spacetime, because physically important quantities such as the ISCO, the light-ring, and the horizon tend to be very close to each other (in the ordinary $r$ coordinate). Now, for most Kerr spacetime related computations it is sufficient to place the computational gridÕs inner boundary at a value  of ${r^*} \approx -100M$ because on a uniformly spaced $r^*$ grid the $r$ coordinate approaches the horizon rather fast: the spacing $\delta r$ becomes smaller than machine precision at modest negative values of $r^*$. However, this is not the case for an NEK spacetime. Therefore, a challenge faced in the context of NEK holes is that the inner boundary must be located at a much larger negative value. This, of course, makes the scale of the computation proportionately larger and therefore, very challenging. A better alternative could be to add another compactifying layer, at large negative values of $r^*$. We did not pursue that approach in this work. Instead, we extended the inner boundary to ${r^*} = -1000M$, which allowed us to consider NEK black holes down to $\epsilon=10^{-6}$. 

As indicated earlier, we only model the plunge and ringdown phases in this current work. We ignored the inspiral phase for several reasons: (i) our focus is the unique nature of the NEK quasi-normal modes, and their associated detectability by current and future gravitational wave observatories, (ii) the inspiral phase waveform has already been analyzed well in detail by other authors \cite{Gralla_2016}, (iii) as pointed out before, computationally modeling the near horizon NEK spacetime was fairly challenging already, and adding the many cycles of the inspiral phase would push the problem beyond current reach without much benefit for the question of interest, and (iv) the inclusion of radiation reaction during the inspiral phase for NEK spacetime requires special treatment \cite{Gralla_2016}, which will only little change the results for the plunge phase and the enduing quasi-normal ringdown.  

\section{Results}\label{sec3}

\subsection{The waveform}

\begin{figure}
\includegraphics[width=8.5cm]{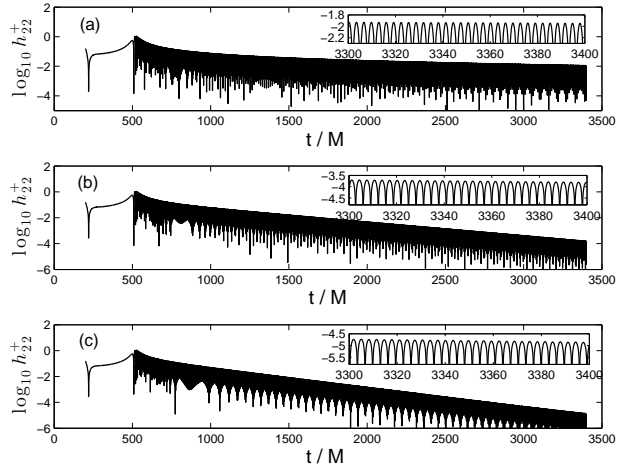}
\caption{The waveforms $h_{22}^+$ as functions of the time $t$ for $\epsilon=1\times 10^{-6}$ (top panel, (a)), for $\epsilon=5\times 10^{-5}$ (middle panel, (b)), and for $\epsilon=1\times 10^{-4}$ (bottom panel, (c)). The insets show short segments of the waveforms to emphasize the oscillations. }
\label{WF}
\end{figure}

We present in Fig.~\ref{WF} the gravitational waveforms $h_{22}^+$ for three values of the spin parameter $\epsilon$. Figure \ref{WF} shows different behavior for small and large values of $\epsilon$. Specifically, for large values of $\epsilon$ the decay of the field is exponential, but for small values of $\epsilon$ it is a power law. Similar results are also found for $h_{22}^{\times}$.

\subsection{The frequency}

\begin{figure}
\includegraphics[width=8.5cm]{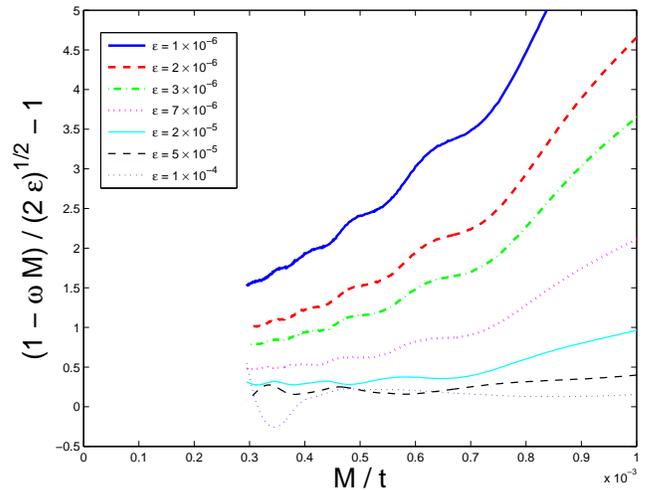}
\caption{The frequency of the gravitational waves as a function of (inverse) time. Specifically, we show $(1-M\omega)/\sqrt{2\epsilon}-1$ as a function of $M/t$. We show the frequency for several values of $\epsilon$.}
\label{freq}
\end{figure}

The frequency of of gravitational radiation is predicted by Gralla {\it et al} \cite{Gralla_2016} to be twice the horizon frequency. The latter is $\Omega_+=a/(2Mr_+)$, where $r_+$ is the event horizon, located at $r_+=M+\sqrt{M^2-a^2}$. To test how our frequencies agree with this prediction, we expand the horizon frequency in $\epsilon$, $1-2M\Omega_+=\sqrt{2\epsilon}+O(\epsilon )$. In Fig.~\ref{freq} we show the angular frequency of the gravitational radiation, $\omega$, as a function of  time for several values of $\epsilon$. Specifically, we show $(1-M\omega )/\sqrt{2\epsilon}-1$ as a function of $M/t$. This quantity should approach 0 as $M/t\to 0$. (More accurately, it should approach $-\epsilon$.) Our data are clearly consistent with this expectation. Notice that for large values of $\epsilon$ (i.e., $\epsilon\gtrsim 10^{-4}$) the frequency starts oscillating at large values of the time. The reason for this behavior is that as the field's amplitude decays exponentially in time the numerical accuracy is compromised. We infer that indeed the frequency of the gravitational waves in the ringdown phase is twice the horizon frequency. 

\subsection{The amplitude}

\begin{figure}
\includegraphics[width=8.5cm]{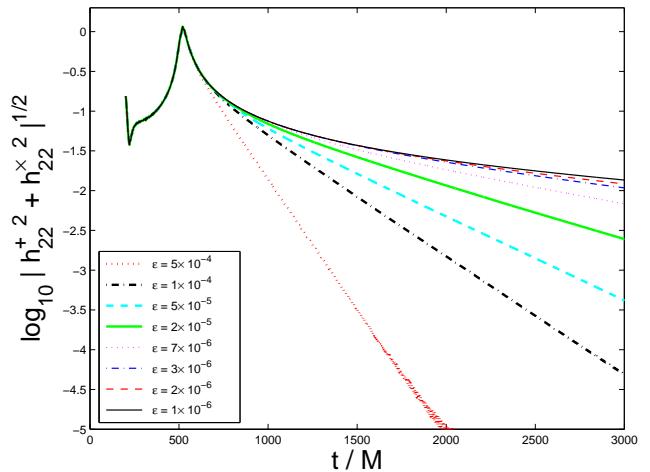}
\caption{The amplitude of the field $h_{22}$ as a function of the time $t$. Shown are the amplitudes for several values of $\epsilon$. In all cases we removed the fields' oscillations to show the amplitude in a clearer way. The actual fields oscillate as shown in Fig.~\ref{WF}.}
\label{amp}
\end{figure}

Next, we present the amplitude of the field in Fig.~\ref{amp}. Specifically, we present $\left|(h_{22}^+)^2+(h_{22}^{\times})^2\right|^{1/2}$ as a function of the time. 

\begin{figure}
\includegraphics[width=8.5cm]{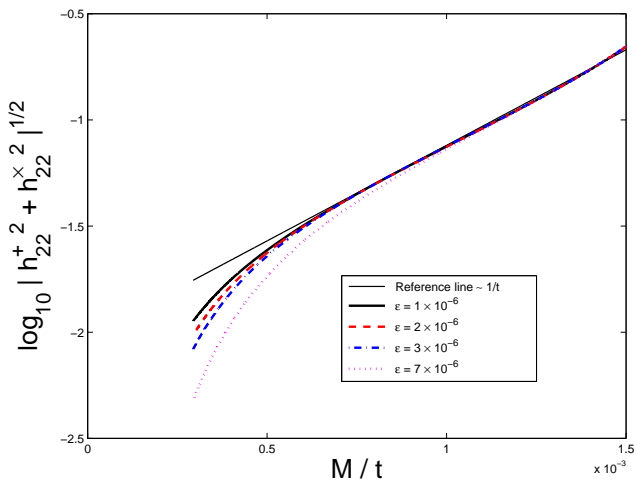}
\caption{The amplitude of the field $h_{22}$ as a function of the time $t$. Shown are the amplitudes for several values of $\epsilon$. In all cases we removed the fields' oscillations to show the amplitude in a clearer way. The actual fields oscillate as shown in Fig.~\ref{WF}.}
\label{inv_amp}
\end{figure}

Figure \ref{amp} suggests a transient behavior of decay rate $~M/t$. To better illustrate this transient behavior we plot in Fig.~\ref{inv_amp} the amplitude as function of inverse time, in addition to a reference line $~M/t$. Figure \ref{inv_amp} shows that the smaller $\epsilon$, the later the amplitude ``peels off'' the $M/t$ behavior, and eventually becomes exponential, as is the case with quasi-normal radiation of non-NEK black holes. This transient behavior is similar to the one found by Yang {\it et al} for a source-free scalar field \cite{Yang-2013}. 

\begin{figure}
\includegraphics[width=8.5cm]{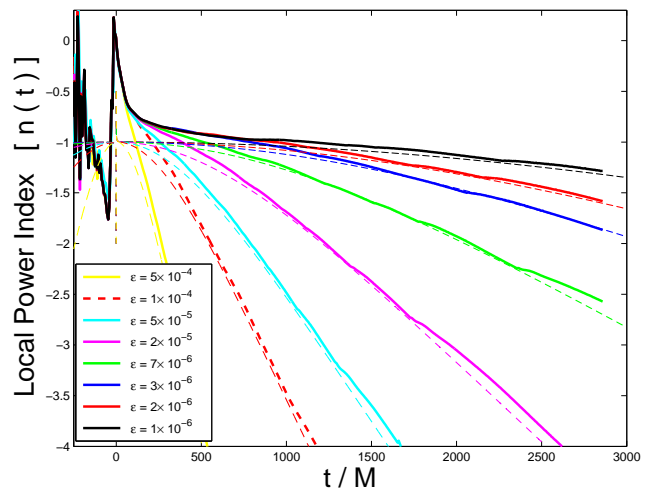}
\caption{The local power indices $n$ of the field's amplitudes as functions of the time $t$ for several values of $\epsilon$. Shown are the $n$ values for the results of the numerical simulations (thick curves) and for the Ansatz (\ref{eq}) (thin dashed curves). At fixed large values of the time the value of $n$ decreases as $\epsilon$ increases. }
\label{lpi}
\end{figure}

In Fig.~\ref{lpi}o we study the decay rate more precisely. Specifically, we consider as Ansatz for the gravitational case the behavior found in \cite{Yang-2013} for a source-free scalar field, specifically, 
\begin{equation}\tag{4}
h_{22}^{+,\times}\approx \sqrt{\epsilon}\,
\frac{e^{-\sqrt{\epsilon /(8t)}}}
{1-e^{-\sqrt{\epsilon /(2t)}}}\, .
\label{eq}
\end{equation}
A similar relation was found in \cite{Yang-2013} for a source-free scalar field after careful analysis of the quasi-normal overtones. Here, we take this behavior as an Ansatz, which we test against the numerical results. We are primarily interested in the local behavior of the field, and therefore use the local power index \cite{burko-ori-97},
\begin{equation}\tag{5}
n(t)=t\,\frac{\dot{h_{22}}}{h_{22}}\, ,
\label{lpi_eq}
\end{equation}
where $h_{22}$ denotes the amplitude of the field, and an overdot denotes a derivative with respect to time. We apply (\ref{lpi_eq}) both to the numerical results for the amplitude, and for the Ansatz (\ref{eq}). Figure \ref{lpi} shows $n(t)$ as function of the time for several values of $\epsilon$. The transient behavior of $n(t)\approx -1$ can be seen as protracting longer the smaller $\epsilon$. We also find that the Ansatz (\ref{eq}) agrees very well with the numerical simulations. 

\begin{figure}
\includegraphics[width=8.5cm]{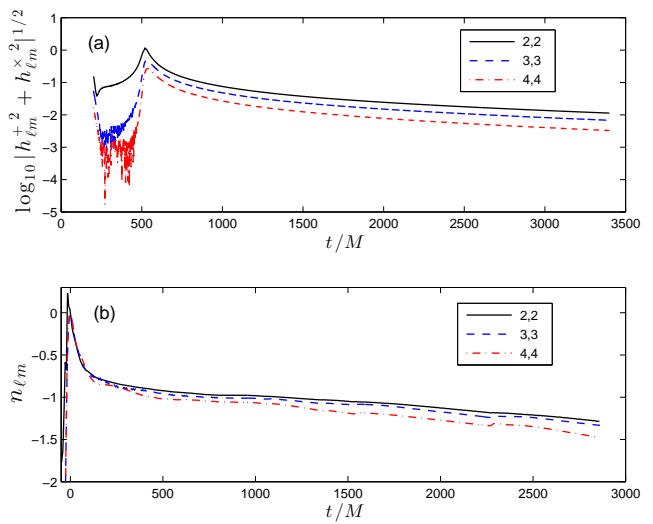}
\caption{The betavior of different $\ell,m$ modes. Upper panel (a): the amplitude as a function of time. Lower panel (b): the local power index as a function of time. In either panel we show the $(2,2)$ mode (solid curve), the $(3,3)$ mode (dashed curve), and the $(4,4)$ mode (dash-dotted curve). For all cases we present the results for $\epsilon=1\times 10^{-6}$. }
\label{ell_m}
\end{figure}

Next, we consider the dependence of the amplitude and the local power index on the mode number.  Specifically, we consider the values $(\ell ,m)=(2,2), (3,3)$ and $(4,4)$. In Fig.~\ref{ell_m} we show the 
amplitude of the field (upper panel) and the local power index  (lower panel) for each mode for $\epsilon=1\times 10^{-6}$. We find that the local power indices are nearly the same, including the transient epoch with decay rate that is inverse in time. The asymptotic amplitudes of the fields are very different, with the $(2,2)$-mode amplitude greater than the $(3,3)$-mode amplitude by a factor $\sim 3$, and the $(3,3)$-mode amplitude greater than the $(4,4)$-mode amplitude by a factor $\sim 5$.

\subsection{Energy flux}

\begin{figure}
\includegraphics[width=8.5cm]{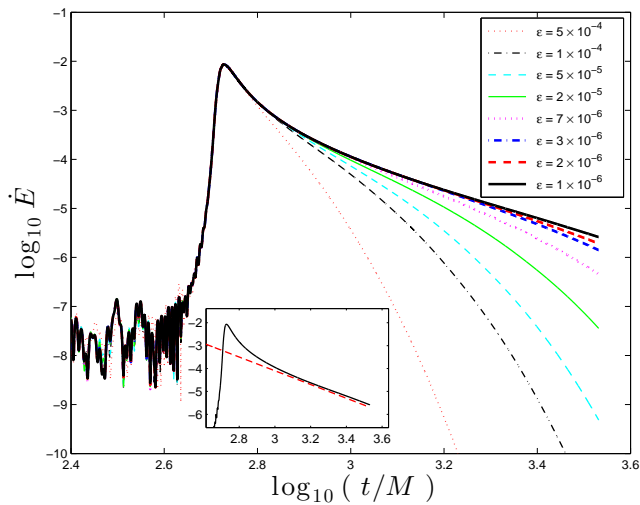}
\caption{The flux of energy at infinity for the sum over all multipoles $\ell$ for the mode $m=2$. Shown are the curves for several values of $\epsilon$. The inset repeats the curves for $\epsilon=1\times 10^{-6}$ and adds the reference line $\sim t^{-3}$. }
\label{flux}
\end{figure}

The transient behavior we show implies that the energy flux too decays at a slower rate. We show in Fig.~\ref{flux} the energy flux for the sum over all multipoles $\ell$ for the mode $m=2$, calculated numerically by 
\begin{equation}\tag{6}
\dot{E}(t)=\frac{1}{4\pi}\;\lim_{r\to\infty} \int  d\Omega\, r^2 \int^t \,dt' \;\psi_4^*(t')\,\psi_4(t')\, ,
\end{equation}
where $\psi_4$ is the Weyl scalar, and $\,d\Omega$ is the line element on the unit 2-sphere. For small values of $\epsilon$ the energy flux at infinity have a transient behavior of $\sim t^{-3}$, and the larger $\epsilon$, the sooner the energy flux deviates from this behavior and becomes exponential, as it is for non-NEK black holes. 

\begin{figure}
\includegraphics[width=8.5cm]{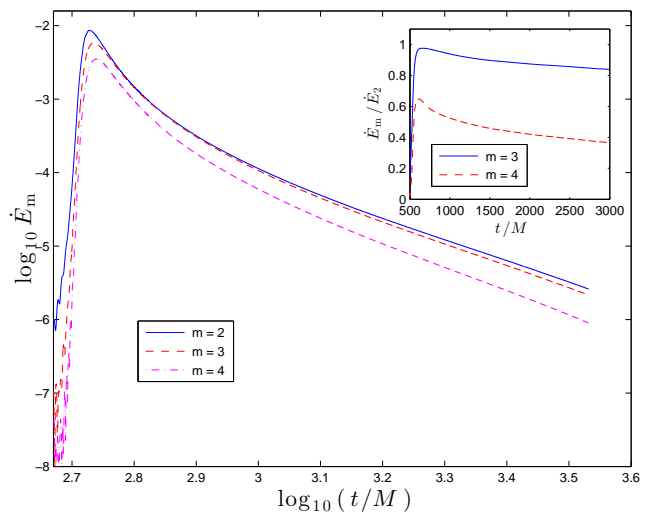}
\caption{The logarithms of the  fluxes of energy at infinity for the sums over all multipoles $\ell$ for the modes $m=2$ (solid curve), $m=3$ (dashed curve), and $m=4$ (dash-dotted curve) as functions of the logarithm of time. The inset shows the ratios 
${\dot E}_m/{\dot E}_2$ for $m=3$ (solid curve) and $m=4$ (dashed curve) as functions of the time. 
For all cases $\epsilon=1\times 10^{-6}$.}
\label{flux_m}
\end{figure}

The relative contributions to the flux at infinity of the different $m$ modes (when all multipoles $\ell$ are summed over) in shown in Fig.~\ref{flux_m}. The contribution of the $m=3$ modes to the flux is comparable to that of the mode $m=2$, at about $85\%$ of the latter during the epoch of inverse time decay. The $m=4$ mode is substantially weaker, at about $40\%$. These results for the plunge are consistent with the results from the earlier inspiral phase: as shown in \cite{Gralla_2016}, the $\ell=2,m=2$ mode contributes only about $10\%$ of the total flux during the near horizon inspiral. 

\section{Detectability}\label{sec4}

As pointed out in \cite{Yang-2013} and in \cite{Gralla_2016} the unique gravitational waveform from an inspiral and plunge into a NEK black holes provides an interesting opportunity for gravitational wave detection. The ``smoking gun" evidence found in \cite{Gralla_2016} will serve as  a unique feature of the gravitational waveform that will facilitate identification. In addition, the slowly decaying quasi-normal modes found in \cite{Yang-2013} will allow for a longer detection of the signal. Here, we apply the ideas that were brought in \cite{Yang-2013} in the context of a source-free scalar field for  gravitational waves. 

Specifically, we consider, for the sake of comparison, a NEK black hole of the same mass as the final mass of the resulting black hole of GW150914, i.e., $M=67\pm 4\; M_{\odot}$ (in the detector frame) \cite{GW150914}. We take however the spin rate of the black hole to be $a/M=1-\epsilon$, where for specificity we take $\epsilon=1\times 10^{-6}$. We then assume that some hypothetical NEK black hole source emits radiation whose maximum strain at the detector is the same as that of GW150914. 

\begin{figure}
\includegraphics[width=8.5cm]{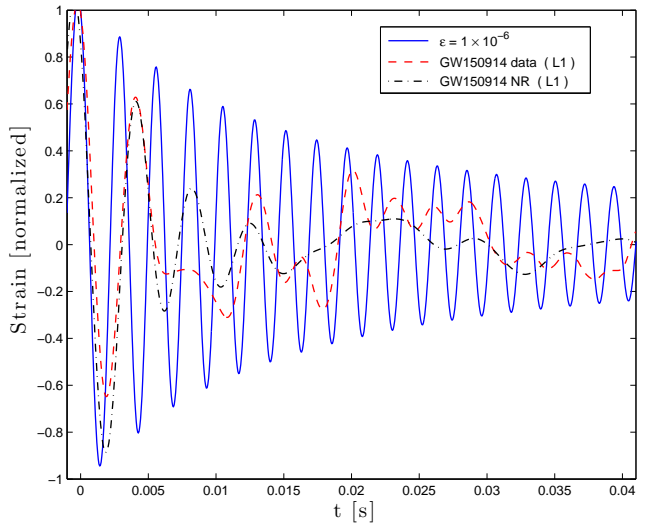}
\caption{The  strain in gravitational waves for a hypothetical NEK source of $\epsilon=1\times 10^{-6}$ that is detected with the same maximal strain at the detector as GW150914, when the mass of the NEK black hole in the detector frame is $M=67\; M_{\odot}$. Shown are the the NEK strain (solid curve), the GW150914 signal in L1 (dashed curve), and the numerical relativity data for GW150914 projected onto the L1 detector in the frequency range 35 -- 350 Hz:  The thin dashed-dotted curve shows a numerical relativity waveform for a system with parameters consistent with those recovered from GW150914 confirmed by an independent calculation}
\label{gw}
\end{figure}

We display in Fig.~\ref{gw} the gravitational waveform from the NEK source superimposed on the signal of GW150914, when both signals are normalized to the same strain amplitude at maximum strain, which we set at $t=0$. The GW150914 data shown are the detected strain in the L1 observatory and the numerical relativity data projected onto the LIGO L1 detector in the frequency band 35 -- 350 Hz \cite{GW150914}. (Similar waveforms were detected also in H1.)

Figure \ref{gw} confirms that indeed the detectability of a NEK black hole is enhanced compared with that of a non-NEK black hole. Specifically, indeed the slow damping (i.e., damping in inverse time instead of exponential) allows for both a higher signal-to-noise ratio and for significantly longer detection times. For higher mass NEK black holes the decay rate is even slower, making the detected signal longer lived. 

Gralla {\it et al} \cite{Gralla_2016} consider the potential confusion of the gravitational waves coming from the near horizon inspiral with the quasi-normal modes. Specifically, in  \cite{Gralla_2016} it is pointed out that the pre-near horizon inspiral history of the gravitational waveforms can remove the ambiguity. We are now in a position to add also the post-near horizon inspiral: the characteristic decay at twice the horizon frequency with decay in inverse time is characteristic of the plunge and initial ringdown radiation from NEK black holes. Depending on the value of $\epsilon$, the observed initial inverse-time decay rate may even change to an exponential decay rate.




\section*{Acknowledgements} 
This research has made use of data, software and/or web tools obtained from the LIGO Open Science Center (https://losc.ligo.org), a service of LIGO Laboratory and the LIGO Scientific Collaboration. LIGO is funded by the U.S. National Science Foundation.

The authors thank Sam Gralla, Niels Warburton, and Scott Hughes for discussions. 

G.K.~acknowledges research support from NSF Grants No.~PHY--1414440 and No.~PHY--1606333, and from the U.S.~Air Force agreement No.~10--RI--CRADA--09.

\end{document}